# *De novo* design of protein binders targeting the human sweet taste receptor as potential sweet proteins

Saisai Ding [a] & Yi Zhang [a]*

[a] Department of Food Science, The Pennsylvania State University, State College, PA 16802, United States

Saisai Ding. E-mail: sqd5856@psu.edu
Yi Zhang, PhD. E-mail: yjz5549@psu.edu

* Indicates corresponding author

**Abstract**

Excessive consumption of dietary sugars is a major contributor to metabolic disorders, driving global interest in finding alternative sweeteners with reduced caloric impact. Natural sweet proteins, such as brazzein, offer exceptional sweetness intensity with little caloric contribution. However, their widespread use is limited by restricted natural diversity, low stability, and high production costs. Recent advances in structural biology and *de novo* protein design provide new opportunities to overcome these limitations through rational engineering. In this study, we report an integrated computational pipeline for the *de novo* design of protein binders targeting the human sweet taste receptor subunit TAS1R2, a key component of the heterodimeric class C G protein-coupled receptor mediating sweetness perception. The workflow combines diffusion-based backbone generation (RFdiffusion), neural network-guided sequence design (ProteinMPNN), structure-based filtering using Boltz-1, and binding energy evaluation via MM/GBSA calculations. Using the recently resolved cryo-EM structure of the TAS1R2 receptor, protein binders were designed to target both the Venus Flytrap Domain and the cysteine-rich domain of TAS1R2. A few designed binders exhibited favorable structural confidence and predicted binding energetics. In particular, Binder2 exhibited brazzein-like structural plausibility through specific short-range CRD contacts, while Binder1 displayed the strongest predicted binding affinity. Structural analyses of the binder-receptor complex revealed distinct binding modes and secondary structure profiles among the designs. This study demonstrates the feasibility of *de novo* designing protein binders that emulate key functional properties of natural sweet proteins, establishing a computational framework for the rational development of next-generation protein-based sweeteners.

# 1. Introduction

Excess dietary sugars, including sucrose and fructose, are major contributors to metabolic diseases such as obesity and type 2 diabetes by promoting excessive caloric intake, hepatic *de novo* lipogenesis, and other mechanisms. There is a growing global effort to reduce added sugar intake and to replace sucrose in foods with high-potency sweeteners (Herman & Birnbaum, 2021; X. Ma et al., 2022; Sievenpiper et al., 2025; Stanhope, 2016; Witek, Wydra, & Filip, 2022). Sweeteners are broadly categorized into synthetic (artificial) and natural compounds. Synthetically produced non-nutritive sweeteners, such as saccharin, aspartame, sucralose, and neotame, are several hundred to several thousand times sweeter than sucrose and can effectively reduce caloric intake when substituted for sugar in foods and beverages. However, their widespread use faces persistent challenges, including undesirable aftertastes, limited consumer acceptance compared with sucrose, and ongoing debates about long-term health consequences (Ahmad, Friel, & Mackay, 2020; Conz, Salmona, & Diomede, 2023; Mbambo, Dlamini, Chukwuma, & Islam, 2020; Suez et al., 2022). For instance, aspartame has been controversially linked to cancer risk, with a large Spanish multicase-control study suggesting possible increased risks for colorectal and stomach cancer among participants with diabetes (Palomar-Cros et al., 2023). Natural sweeteners derived from plant sources include stevia glycosides from *Stevia rebaudiana* and mogrosides from *Siraitia grosvenorii*, which are often accompanied by undesirable aftertastes, most notably the bitter or lingering notes associated with stevia glycosides, as well as the metallic or licorice-like flavors reported for mogroside extracts, restricting their broader application in foods (Al-Taweel, Azzam, Khaled, & Abdel-Aziz, 2021; Chranioti, Chanioti, & Tzia, 2016; Muñoz-Labrador et al., 2021; Orellana-Paucar, 2023).

Sweet proteins are a small family of plant-derived proteins occurring naturally in tropical fruits, including thaumatin (*Thaumatococcus daniellii*), monellin (*Dioscoreophyllum cumminsii*), brazzein (*Pentadiplandra brazzeana*), and neoculin (*Curculigo latifolia*). These proteins elicit sweetness ranging from several hundred to over three thousand times that of sucrose, yet contribute virtually no calories (de Jesús-Pires et al., 2020; Li, Zheng, Yu, Wang, & Pan, 2023; Lifshitz et al., 2025; López-Plaza et al., 2024; Lynch, Wang, Vo, Tafazoli, & Ryder, 2023; M. Ma et al., 2025). Natural sweet proteins often display modest stability and poor tolerance to heat and pH fluctuations, making large-scale production technically demanding and costly. Recent progress has improved the stability and applicability of sweet proteins. Redesign of monellin produced variants with melting temperatures close to 96 °C, which was about 20 °C higher than the native, while retaining sweetness (Lifshitz et al., 2025; Liu, Xu, Ma, You, Ye, & Liu, 2024; M. Ma et al., 2025; Miao et al., 2022; Wang et al., 2025; Zuo et al., 2024). Although these strategies have advanced the stabilization of sweet proteins, broader limitations exist, including the limited natural availability of sweet proteins and the restrictions arising from patent protection. Sweetness perception is mediated primarily by the human sweet taste receptor, a heterodimeric G protein-coupled receptor (GPCR) composed of the TAS1R2 and TAS1R3 subunits, which belong to the class C family of GPCRs. Each subunit contains an extracellular Venus Flytrap Domain (VFTD), a cysteine-rich domain (CRD), and a transmembrane domain (TMD). A recent study showed that upon ligand binding, conformational rearrangements were predominantly observed in the VFTD of TAS1R2, leading to a closure of the orthosteric pocket, whereas the VFTD of TAS1R3 remained largely unchanged (Shi et al., 2025). Although these natural sweet proteins produce similar sweet taste perceptions, little sequence or structural homology has been identified among them.

The convergence of *de novo* protein design algorithms with high-resolution structural biology has created an unprecedented opportunity to engineer novel food proteins. Recent generative frameworks, such as RFdiffusion, ProteinMPNN, and BindCraft, combined with structure-prediction advances exemplified by AlphaFold3 and Boltz, now permit the systematic exploration of *de novo* protein design with atomic-level precision (Abramson et al., 2024; Dauparas et al., 2022; Pacesa et al., 2025; Watson et al., 2023; Wohlwend et al., 2025). These models benefit the *de novo* sweet protein design by implementing a structure-based binder design philosophy: RFdiffusion generates binder conformations compatible with the human sweet protein receptor interface; ProteinMPNN designs sequences that satisfy these geometries; and Bolt-1 filters the resulting complexes using ipTM, confidence_score, and related structural quality metrics. In 2025, the recent cryo-EM resolution of the human TAS1R2–TAS1R3 heterodimer revealed, for the first time, the modular architecture and inter-subunit organization underlying sweetness perception, in which ligand binding induces a marked closure of the TAS1R2 VFTD while the TAS1R3 VFTD remains open. This asymmetry is propagated through the CRD to the transmembrane domains, providing a structural template for rational targeting (Juen et al., 2025; Shi et al., 2025). Together, these developments enable mechanistically guided design of sweet taste receptor binders, laying the groundwork for engineering sweet proteins.

In this study, we establish an integrated *de novo* design pipeline to generate binders of the human sweet taste receptor as potential sweet proteins. Our framework combines generative backbone construction, sequence optimization, structural filtering, and binding energy evaluation. To the best of our knowledge, this work represents the first integrated pipeline that performs *de novo* binder design targeting the human sweet taste receptor, thereby offering a computational framework for the rational creation of novel sweet proteins.

## 2. Materials and methods

### 2.1. TAS1R2 structure preparation

The amino acid sequence of the human taste receptor type 1 member 2 (TAS1R2) was obtained from UniProtKB (accession number Q8TE23) (Consortium, 2019), comprising 839 residues. The three-dimensional structure was derived from the cryo-EM structure of the TAS1R2–TAS1R3 heterodimer (Shi et al., 2025) (PDB ID: 9NOR, chain B) and from the AlphaFold2-predicted model generated using the UniProt sequence Q8TE23.

### 2.2. Binder backbone generation

Binder backbones were generated with the RFdiffusion binder design protocol (https://github.com/RosettaCommons/RFdiffusion) (Watson et al., 2023), using TAS1R2 as the fixed target structure derived from PDB 9NOR chain B and the AlphaFold2 model based on UniProtKB Q8TE23. A single continuous binder chain of 128 amino acids was specified. Hot-spot conditioning was applied for binder design under two schemes. (i) VFTD cavity (Assadi-Porter, Tonelli, Maillet, Markley, & Max, 2010): the hotspot set comprised residues in the VFTD cavity, S40, K65, I67, L71, Y103, D142, N143, S144, E145, S165, A166, I167, S168, A187, H190, Y215, P277, D278, L279, Y282, E302, S303, W304, D307, V309, E382, R383, and V384, provided to RFdiffusion as contact-bias residues to guide binder placement. (ii) CRD region: hotspots were defined as residues belonging to the CRD of TAS1R2 (Juen et al., 2025; Shi et al., 2025). Across the two hotspot definitions, a total of 5,000 candidate 128-residue backbones were sampled.

### 2.3. *Binder sequence design*

Following backbone generation, candidate binders were subjected to sequence design using the ProteinMPNN–FastRelax workflow implemented in the dl_binder_design protocol (https://github.com/nrbennet/dl_binder_design) (Bennett et al., 2023; Dauparas et al., 2022). For each protein–binder complex, the binder sequence was masked and ProteinMPNN was tasked with assigning amino acids to the 128-residue binder backbone. The newly designed sequence was then threaded back onto the backbone within the complex, followed by structural relaxation with PyRosetta FastRelax (Chaudhury, Lyskov, & Gray, 2010). The relaxed complexes were subsequently reintroduced into ProteinMPNN for iterative optimization. Sequence generation was performed using the ProteinMPNN–FastRelax Binder Design module with the parameter setting -seqs_per_struct = 1.

### 2.4. *Binder evaluation*

Designed binder sequences from ProteinMPNN were evaluated using Boltz-1 (Wohlwend et al., 2025), an open-source implementation of AlphaFold3 (Abramson et al., 2024). For each binder–receptor complex, structural predictions were generated with the --use_msa_server option enabled to leverage multiple sequence alignment information. Five independent predictions were obtained for each binder–receptor pair. The resulting complexes were ranked and filtered based on multiple evaluation metrics, including an overall confidence_score > 0.6, and the average predicted local distance difference test score of the complex (complex_pLDDT >0.7).

### 2.5. *Structural preparation and visualization*

The filtered binder–receptor complex structures were further processed using Schrödinger 2023.4 (Madhavi Sastry, Adzhigirey, Day, Annabhimoju, & Sherman, 2013). Hydrogen-bond assignments were automatically optimized, and restrained minimizations were performed, optionally deleting water molecules to improve the quality of the structural models prior to energy evaluation. For visualization, both the designed binders and the final binder–receptor complexes were rendered in ChimeraX 1.7 (Meng et al., 2023) and Pymol 3.0.0 (DeLano, 2002).

### 2.6. *Binding energy calculation*

For the top five binder–receptor complexes selected from the Boltz-1 predictions, binding free energies were estimated using the Uni-GBSA workflow (Yang, Bo, Xu, Xu, Wang, & Zheng, 2023), an automated pipeline for MM/GB(PB)SA calculations. Each complex structure was processed with Uni-GBSA under default settings, and the GBSA values were used as quantitative measures of predicted complex binding affinity.

### 2.7. *Reference sweet proteins*

For comparative evaluation, the representative sweet proteins were included as benchmarks. Brazzein (Ming & Hellekant, 1994), a naturally occurring sweet protein originally isolated from *Pentadiplandra brazzeana*, was used. The sequence is: DKCKKVYENYPVSKCQLANQCNYDCKLDKHARSGECFYDEKRNLQCICDYCEY.

*2.8. Statistical analysis and visualization*

Statistical analyses were conducted in Python using pandas (McKinney, 2011), NumPy (Van Der Walt, Colbert, & Varoquaux, 2011), SciPy (Virtanen et al., 2020), Matplotlib (Hunter, 2007), and Pingouin. Group differences were first assessed with Welch's ANOVA, which accommodates unequal variances and sample sizes. Post hoc comparisons were performed using the Games–Howell test, and results were summarized with compact letter displays (CLDs) to indicate groups not significantly different at $\alpha = 0.05$.

## 3. Results and Discussion

### 3.1. Design strategy for TAS1R2 binders

The sweet taste receptor subunit TAS1R2 belongs to the class C family of GPCRs and has three modular domains: the extracellular Venus Flytrap Domain (VFTD), a Cysteine-Rich Domain (CRD), and the seven-transmembrane (7TM) region (**Fig. 7a**). Previous studies have implicated that the TAS1R2 subunit is a key determinant of receptor activation (Assadi-Porter et al., 2010). The VFTD is identified as the site for ligand recognition, including small-molecule sweeteners such as sucralose (Shi et al., 2025) and glucose (Assadi-Porter et al., 2010), as well as sweet proteins like brazzein (Assadi-Porter et al., 2010). The CRD region has also been reported in mediating interactions with sweet proteins such as brazzein (Assadi-Porter et al., 2010). Based on these observations, our design strategy broadened the binder search space and focused on both the VFTD and CRD of TAS1R2 as regions of interest.

Using recent methodological advances, we constructed a design-and-screen pipeline that integrates generative design with hierarchical structural screening (**Fig. 7b**). In the first stage, we generated *de novo* binder backbone using RFdiffusion, which is a diffusion-based framework that produces diverse structural scaffolds and thereby enables broad sampling of conformational space for candidate binders (Watson et al., 2023). In the second stage, sequence design was carried out with ProteinMPNN, which maps the generated backbones to amino acid sequences that are physically realistic and compatible with the targeted receptor interface (Dauparas et al., 2022). In the third stage, structural filtering was performed using Boltz-1 (Wohlwend et al., 2025), a structure prediction model that evaluates binder–receptor interactions and eliminates unstable or misfolded candidates. Finally, in the fourth stage, binding energy evaluation was conducted using the Uni-GBSA workflow (Yang et al., 2023). The resulting GBSA binding free energies were then used as quantitative criteria to prioritize designs predicted to achieve favorable interaction energetics (Valdés-Tresanco, Valdés-Tresanco, Valiente, & Moreno, 2021).

## 3.2. Evaluation of designed TAS1R2 binders by Boltz-1

To benchmark the structural plausibility of the designed TAS1R2 binders, we applied Boltz-1 to predict receptor–binder complexes and extracted multiple confidence metrics (**Fig. 8**). For the ipTM (**Fig. 8a**), brazzein and Binder2 exhibited the highest values between 0.5 and 0.6, significantly outperforming other designed binders (Binder1, Binder3, Binder4; $p<0.05$). Binder5 displayed intermediate ipTM scores, suggesting moderate but non-negligible interface quality. For the complex predicted distance error (complex_pde) (**Fig. 8b**), brazzein and Binder2 were significantly lower than other binders ($p<0.05$), among which Binder3 achieved the highest values (>1.1), followed by Binder1, Binder4, and Binder5. This suggests that Binder2 has stable docking orientations. The complex pLDDT scores (**Fig. 8c**) were consistently high (>0.78) across all complexes, confirming good overall model confidence. Notably, brazzein showed the highest pLDDT (0.84), while Binder5 was significantly lower compared to other designs ($p<0.05$). The overall confidence score (**Fig. 8d**) paralleled the ipTM trend: brazzein and Binder2 showed the strongest signals (>0.75), significantly higher than Binder1, Binder3, and Binder4 ($p<0.05$). The ipTM is as a robust binary predictor of binding competence, effectively distinguishing receptor-compatible complexes from non-binding configurations. In summary, Binder2 most closely matches the performance of the native sweet protein brazzein among the designed binders, particularly with respect to ipTM and overall confidence score, suggesting a higher likelihood of forming a stable TAS1R2-binding interface.

## 3.3. Binding poses of designed TAS1R2 binders

To visualize the binding poses of the designed binders, complexes predicted by Boltz-1 were depicted in **Fig. 9**. From left to right, the complex structures correspond to Binder1 to Binder5 with the receptor. The predicted binding sites varied across designs. Binder1 and Binder2 docked primarily to the VFTD, consistent with the known sweetener-binding pocket. Binder3 and Binder4 are associated with the CRD, while Binder5 was localized to a distal region of the VFTD.

Secondary structure analysis of the designed binders revealed distinct, binder-specific structural profiles (**Fig. 10**). Binder1 contained ~56% helix, 24% strand, and 20% coil, representing a balanced fold. Binder2, which displayed the most favorable binding energetics, had 92% helix and 8% coil, strongly helix-dominated, indicating a highly ordered architecture. Binder3

Page 8 of 23

also showed a helical preference with 78% helix, 0% strand and 22% coil but localized to the CRD. Binder4 exhibited a mixed composition with 58% helix, 26% strand, and 16% coil, while Binder5 was largely helical (85% helix, 0% strand, 15% coil). It is worth noting that the generative design pipeline, as RFdiffusion combined with ProteinMPNN tends to favor confident sampling of helical backbones, making helices more readily designed than extended strand-rich architectures.

### 3.4. Binding energy evaluation

We next quantified the binding energetics of natural sweet protein brazzein and the designed binders to the receptor using the Uni-GBSA workflow (**Fig. 11**). Brazzein exhibited a binding free energy of −70.7 kcal·mol$^{-1}$, serving as a reference benchmark. Binder1 displayed an exceptionally favorable energy of −115.6 kcal·mol$^{-1}$, which was markedly stronger than that of brazzein and all other designs. Binder2 showed a moderately favorable binding energy (−62.6 kcal·mol$^{-1}$), comparable to brazzein. The structural basis for Binder2 binding was further investigated at the TAS1R2 CRD domain (**Fig. 12**). Binder2 established specific short-range contacts within 3.0 Å, in which ARG37 and THR55 from the binder interact with GLY509 and ALA228 of the receptor, respectively. In contrast, Binder3, Binder4, and Binder5 showed substantially lower energies (−43.7 to −48.4 kcal·mol$^{-1}$), suggesting less stable receptor interactions.

### 3.5. Discussion

Sweet proteins represent a unique class of natural sweeteners with considerable potential for food industry applications. Their extraordinary sweetening potency allows for effective use at low concentrations. Compared with conventional carbohydrate-based sweeteners, such as sucrose, sweet proteins provide the additional benefit of negligible caloric contribution. This makes them particularly attractive alternatives in sugar-reduction strategies aimed at addressing obesity, diabetes, and other diet-associated health concerns. Considerable efforts have therefore been devoted to improving natural sweet proteins' physicochemical properties, especially stability, solubility, and sweetness retention. Recent studies have reported progress through targeted modifications, including chimeric redesign of monellin and mutational strategies guided by structural analysis and electrostatic surface potentials (Lucignano et al., 2024; Yasui, Nakamura, & Yamashita, 2021). While such approaches have improved the thermal stability and broadened the pH tolerance of natural sweet proteins, further advances need to be made for industrial production and application.

We applied an integrated computational design and evaluation pipeline to generate and assess binders for TAS1R2 receptor. Regarding sweet taste receptor, structural understanding of the cognate receptor, TAS1R2–TAS1R3 heterodimer, was incomplete before year 2025, which was one limitation that hindered rational design of sweet proteins that can mimic the nuanced interaction modes of small-molecule sweeteners. The binder design in this study was based on the recently-reported cryo-EM structures of the TAS1R2–TAS1R3 heterodimer (Juen et al., 2025; Shi et al., 2025), providing informative molecular insights into receptor architecture and ligand-binding domains. Moreover, recent advances in *de novo* protein design frameworks, such as diffusion-based backbone generation and neural network-driven sequence design, have markedly expanded the structural and functional search space. Meanwhile, the advances in protein dynamics simulations have revealed the conformational ensembles associated with rare functional states. Collectively, these developments make it feasible to design protein-based binders that emulate, or potentially surpass, the functional properties of natural sweet proteins.

By combining RFdiffusion-based scaffold generation, ProteinMPNN sequence design, structural screening with Boltz-1, and binding free energy evaluation using the Uni-GBSA workflow, we systematically prioritized designs predicted to achieve favorable interaction energetics. Among the generated binders, Binder1 displayed the strongest predicted binding affinity, while Binder2 emerged as the most structurally plausible binder, closely approaching brazzein in both predicted confidence metrics and receptor engagement. Structural analyses highlighted specific short-range contacts between Binder2 and the CRD of TAS1R2, providing a mechanistic rationale for its intermediate binding energetics relative to brazzein. These results demonstrate the feasibility of *de novo* design of binders as potential sweet proteins. A design pipeline was also established to integrate state-of-the-art generative modeling, structural prediction, and energetic evaluation.

It is possible that a more rational design of binders can be obtained based on the experimentally resolved heterodimeric structures of TAS1R2–TAS1R3 receptors. Moreover, a deeper understanding of the neurobiological mechanisms of sweetness perception, including crosstalk with other chemosensory receptors and central processing pathways, will provide a broader context for tailoring protein-based sweeteners.

## 4. Conclusion

We established a computational framework for the *de novo* design of binders, as potential sweet proteins, targeting the human sweet taste receptor TAS1R2, by integrating generative modeling, structural prediction, and binding energy evaluation. The designed binders demonstrated varying degrees of structural plausibility and interaction energetics relative to the natural sweet protein brazzein, with Binder2 and Binder1 being the most promising candidates. This work illustrates the feasibility of designing *de novo* sweet proteins through modern design pipelines. Future work could focus on the synthesis of Binder1 and Binder2 for *in vitro* evaluation of sweetness.

**Acknowledgement**

This work was supported by Rasmussen Endowment by The Pennsylvania State University. We also thank Pennsylvania State University's Roar Collab supercomputer for their hardware.

**Figures Captions**

**Fig. 1.** Structural basis and computational design pipeline for TAS1R2 binders. **(a)** Domain organization of the TAS1R2 subunit of the sweet taste receptor. **(b)** Overview of the computational design-and-screen pipeline.

**Fig. 2.** Bolt-1 evaluation of designed TAS1R2 binders in comparison with the natural sweet protein brazzein. **(a)** ipTM of each binder–receptor complex. **(b)** Complex predicted distance error (complex_pde). **(c)** Average complex pLDDT, reflecting overall structural confidence. **(d)** Overall Boltz-1 confidence score.

**Fig. 3.** Binding poses of designed TAS1R2 binders. TAS1R2 is shown in purple. **(a)** Binder1 (pink). **(b)** Binder2 (navy). **(c)** Binder3 (cyan). **(d)** Binder4 (yellow). **(e)** Binder5 (orange). **(f)** Brazzein (light blue).

**Fig. 4.** Secondary structure composition of designed binders. Fractions of helix (red), strand (blue), and coil (green) are shown for each designed binder.

**Fig. 5.** Predicted binding free energies of brazzein and designed TAS1R2 binders.

**Fig. 6.** Binder2 (blue) interaction with the TAS1R2 receptor (purple). A zoomed-in view highlights the interfacial interactions: residues ARG37 and THR55 in Binder2 (yellow sticks) form close contacts with residues GLY509 and ALA228 (cyan sticks) within the CRD of TAS1R2.

**Figure 1**

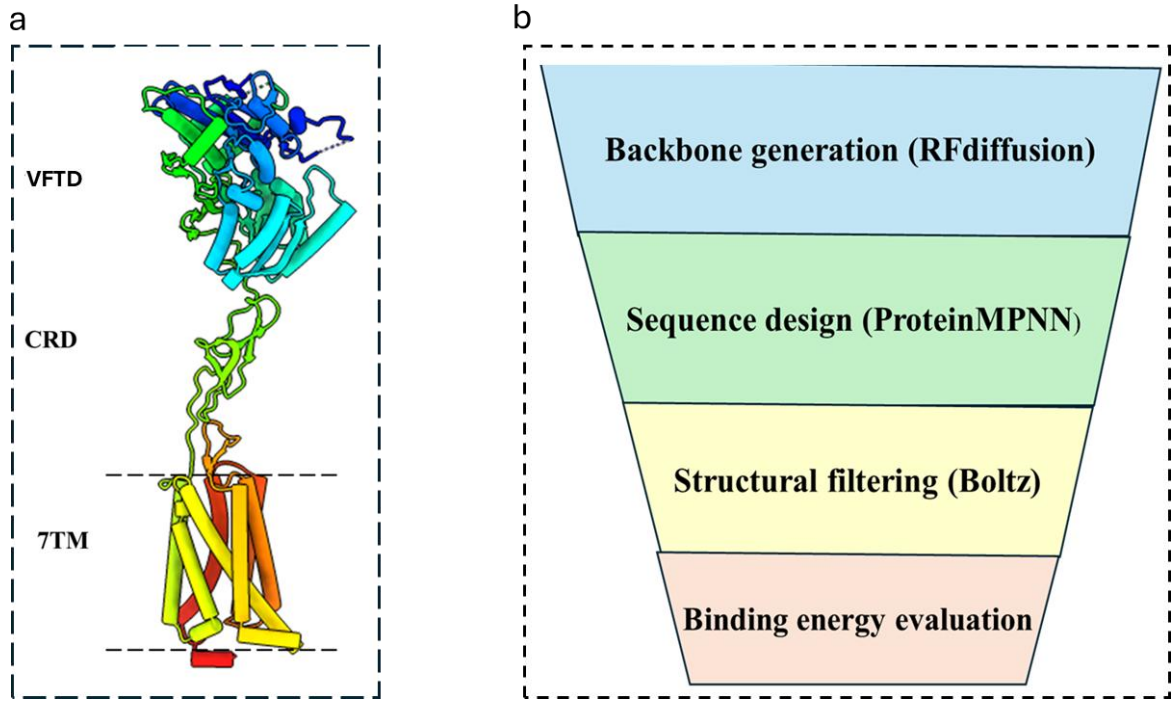

**Fig. 7.** Structural basis and computational design pipeline for TAS1R2 binders. **(a)** Domain organization of the TAS1R2 subunit of the sweet taste receptor. **(b)** Overview of the computational design-and-screen pipeline.

**Figure 2**

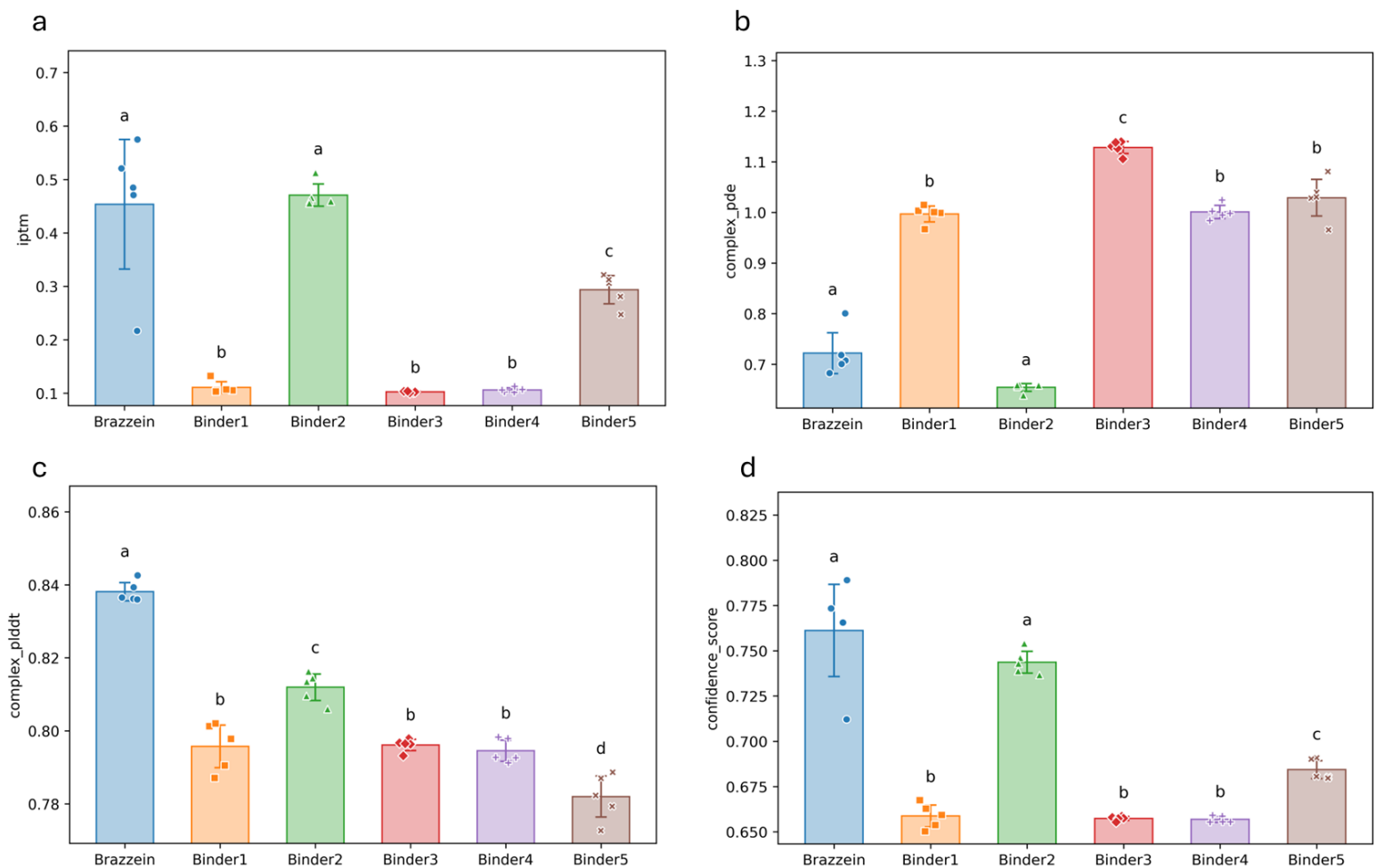

**Fig. 8.** Bolt-1 evaluation of designed TAS1R2 binders in comparison with the natural sweet protein brazzein. **(a)** ipTM of each binder–receptor complex. **(b)** Complex predicted distance error (complex_pde). **(c)** Average complex pLDDT, reflecting overall structural confidence. **(d)** Overall Boltz-1 confidence score.

Figure 3

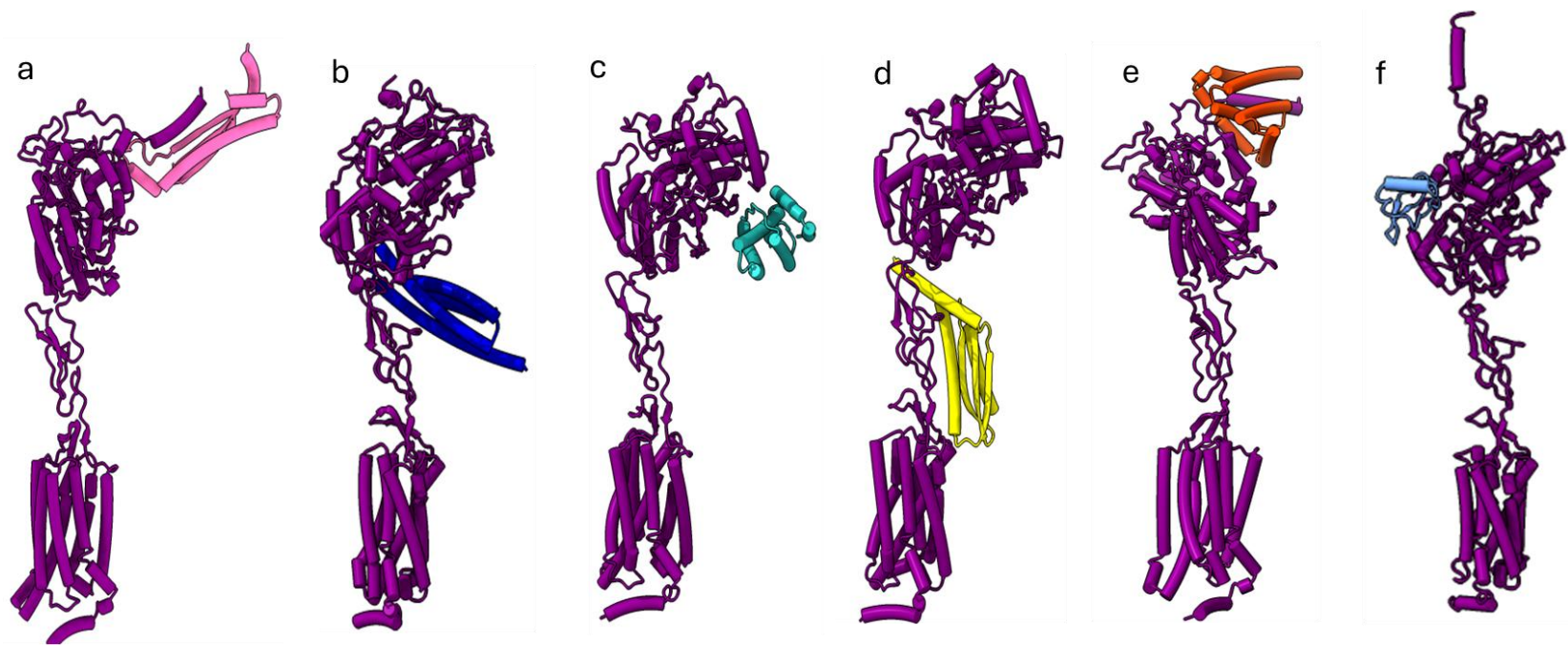

**Fig. 9.** Binding poses of designed TAS1R2 binders. TAS1R2 is shown in purple. **(a)** Binder1 (pink). **(b)** Binder2 (navy). **(c)** Binder3 (cyan). **(d)** Binder4 (yellow). **(e)** Binder5 (orange). **(f)** Brazzein (light blue).

**Figure 4**

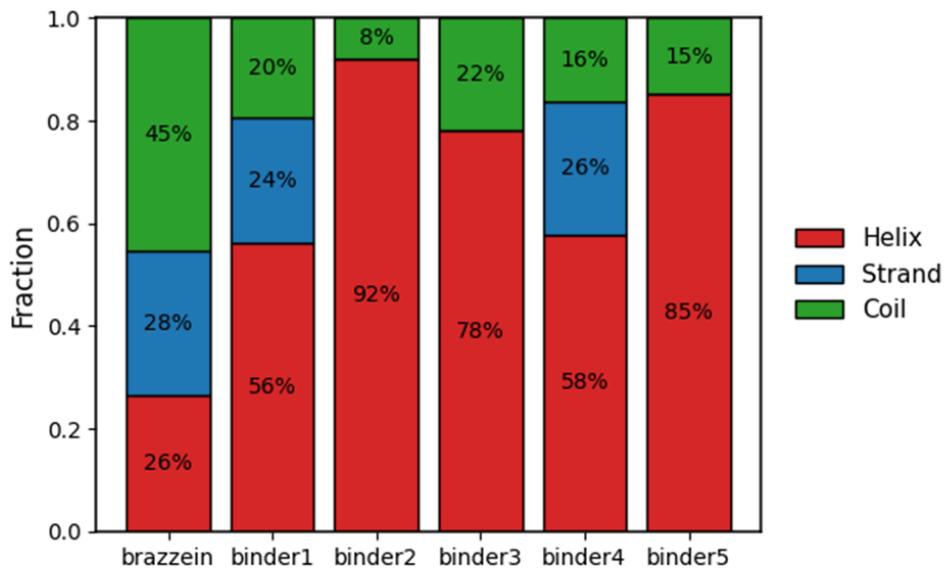

**Fig. 10.** Secondary structure composition of designed binders. Fractions of helix (red), strand (blue), and coil (green) are shown for each designed binder.

**Figure 5**

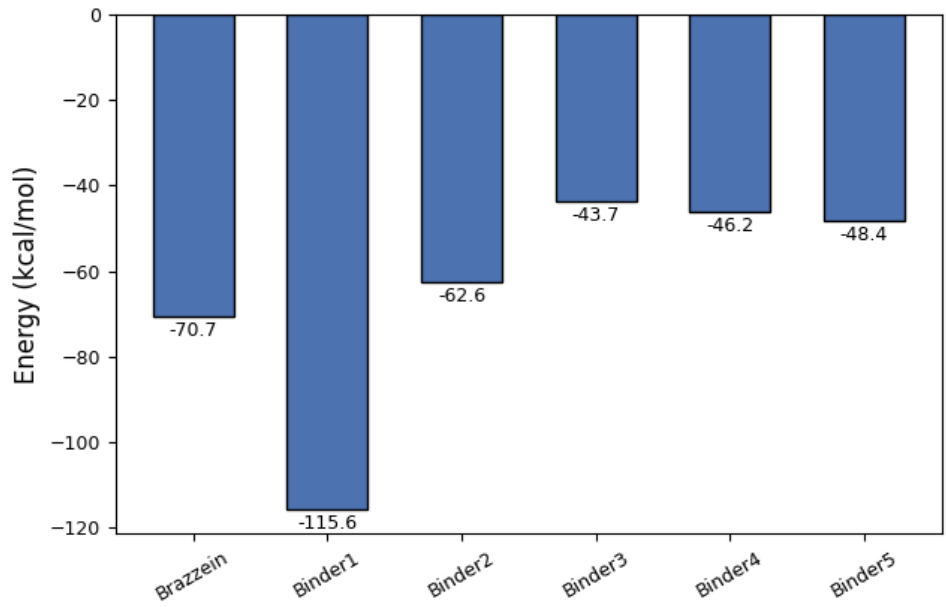

**Fig. 11.** Predicted binding free energies of brazzein and designed TAS1R2 binders.

**Figure 6**

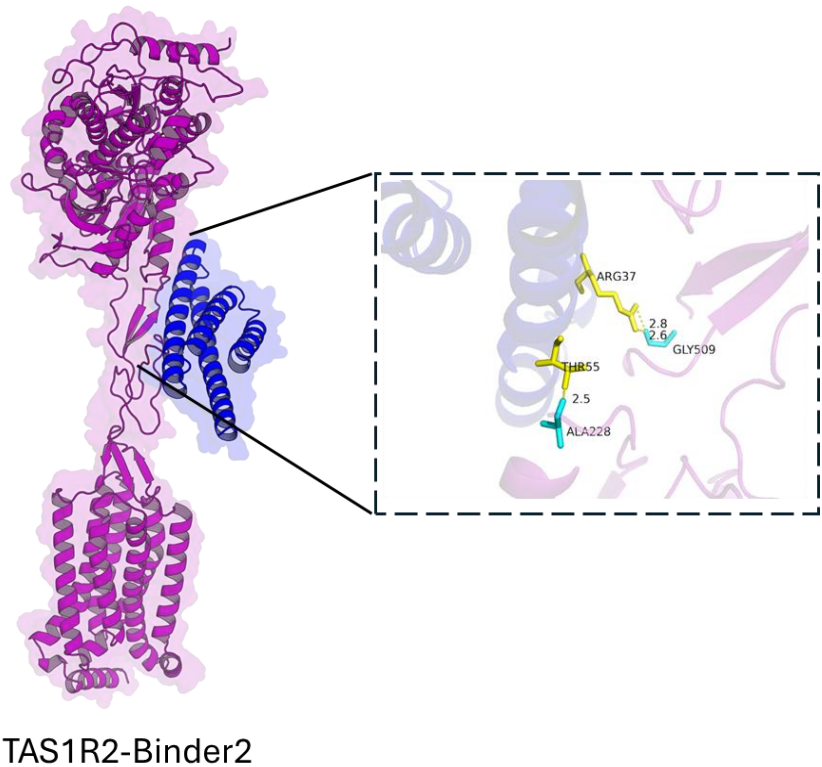

TAS1R2-Binder2

**Fig. 12.** Binder2 (blue) interaction with the TAS1R2 receptor (purple). A zoomed-in view highlights the interfacial interactions: residues ARG37 and THR55 in Binder2 (yellow sticks) form close contacts with residues GLY509 and ALA228 (cyan sticks) within the CRD of TAS1R2.

de Jesús-Pires, C., Ferreira-Neto, J. R., Pacifico Bezerra-Neto, J., Kido, E. A., de Oliveira Silva, R. L., Pandolfi, V., . . . Pio-Ribeiro, G. (2020). Plant thaumatin-like proteins: function, evolution and biotechnological applications. *Current Protein and Peptide Science, 21*(1), 36-51.

DeLano, W. L. (2002). Pymol: An open-source molecular graphics tool. *CCP4 Newsl. protein crystallogr, 40*(1), 82-92.

Herman, M. A., & Birnbaum, M. J. (2021). Molecular aspects of fructose metabolism and metabolic disease. *Cell metabolism, 33*(12), 2329-2354.

Hunter, J. D. (2007). Matplotlib: A 2D graphics environment. *Computing in science & engineering, 9*(03), 90-95.

Juen, Z., Lu, Z., Yu, R., Chang, A. N., Wang, B., Fitzpatrick, A. W., & Zuker, C. S. (2025). The structure of human sweetness. *Cell*.

Li, K., Zheng, J., Yu, L., Wang, B., & Pan, L. (2023). Exploration of the strategy for improving the expression of heterologous sweet protein monellin in Aspergillus niger. *Journal of Fungi, 9*(5), 528.

Lifshitz, Y., Paz, S., Saban, R., Zuker, I., Shmuely, H., Gorshkov, K., . . . Amiram, G. (2025). Safety Evaluation of Serendipity Berry Sweet Protein From Komagataella phaffii. *Journal of Applied Toxicology*.

Liu, Y., Xu, J., Ma, M., You, T., Ye, S., & Liu, S. (2024). Computational design towards a boiling-resistant single-chain sweet protein monellin. *Food Chemistry, 440*, 138279.

López-Plaza, B., Álvarez-Mercado, A. I., Arcos-Castellanos, L., Plaza-Diaz, J., Ruiz-Ojeda, F. J., Brandimonte-Hernández, M., . . . Palma-Milla, S. (2024). Efficacy and Safety of Habitual Consumption of a Food Supplement Containing Miraculin in Malnourished Cancer Patients: The CLINMIR Pilot Study. *Nutrients, 16*(12), 1905.

Lucignano, R., Spadaccini, R., Merlino, A., Ami, D., Natalello, A., Ferraro, G., & Picone, D. (2024). Structural insights and aggregation propensity of a super-stable monellin mutant: A new potential building block for protein-based nanostructured materials. *International Journal of Biological Macromolecules, 254*, 127775.

Lynch, B., Wang, T., Vo, T., Tafazoli, S., & Ryder, J. (2023). Safety evaluation of oubli fruit sweet protein (brazzein) derived from K omagataella phaffii, intended for use as a sweetener in food and beverages. *Toxicology research and application, 7*, 23978473231151258.Page 20 of 23